\documentclass{article}

\usepackage{amssymb,amsfonts,amsmath}
\usepackage{cite,enumerate,float}
\usepackage{color}
\usepackage{tikz}
\usetikzlibrary{arrows,snakes,backgrounds}
\usepackage{url}
\usepackage[vcentermath]{youngtab}

\def\be{\begin{eqnarray}}
\def\ee{\end{eqnarray}}
\def\nn{\nonumber}

\def\p{\partial}

\def\wc{weak composition\ }
\def\wcs{weak compositions\ }

\newcommand{\ttop}[1]{
  q^{\hat{D}_#1}
}

\definecolor{red}{rgb}{1,0,0}
\definecolor{orange}{rgb}{1,0.5,0}
\definecolor{violet}{rgb}{0.7,0,1}

\hoffset= - 1.0in         

\begin{document}

\title{\vspace{1.cm}\bf
Generating twisted Cherednik eigenfunctions
}

\author{
A. Mironov$^{b,c,d,}$\footnote{mironov@lpi.ru,mironov@itep.ru},
A. Morozov$^{a,c,d,}$\footnote{morozov@itep.ru},
A. Popolitov$^{a,c,d,}$\footnote{popolit@gmail.com}
}

\date{ }

\maketitle

\vspace{-6cm}

\begin{center}
FIAN/TD-06/26  \hfill {\bf to the memory of}\\
ITEP/TH-09/26   \hfill {\bf Francesco Calogero}\\
IITP/TH-09/26  \hfill \phantom{.}\\
MIPT/TH-09/26 \hfill \phantom{.}
\end{center}

\vspace{4.cm}

\begin{center}
$^a$ {\small {\it MIPT, Dolgoprudny, 141701, Russia}}\\
$^b$ {\small {\it Lebedev Physics Institute, Moscow 119991, Russia}}\\
$^c$ {\small {\it NRC ``Kurchatov Institute", 123182, Moscow, Russia}}\\
$^d$ {\small {\it Institute for Information Transmission Problems, Moscow 127994, Russia}}
\end{center}

\vspace{.1cm}

\begin{abstract}
Hamiltonians ${\cal H}^{a}_k$ of new integrable systems associated with the integer rays $(-1,a)$ (commutative subalgebras) of Ding-Iohara-Miki (DIM) algebra in the $N$-body representation are closely related to commuting twisted Cherednik Hamiltonians $\mathfrak{C}_i^{(a)}$, ${\cal H}^{a}_k = \sum_{i=1}^N  (\mathfrak{C}_i^{(a)})^k$.
Moreover, symmetric combinations of eigenfunctions in the twisted Cherednik system were recently shown
to produce the DIM Hamiltonian eigenstates.
We explicitly construct these twisted Cherednik eigenfunctions recurrently by action of some
(creation and permutation) operations.
It resembles of a far-going generalization of Kirillov-Noumi operators,
but exact relation remains to be specified.
\end{abstract}

\bigskip

\newcommand\smallpar[1]{
  \noindent $\bullet$ \textbf{#1}
}

\section{Introduction}

Many-body integrable systems of the Calogero-Moser-Sutherland-Ruijsenaars-Schneider type \cite{Cal,CS,RS} attract attention already during more than half a century, since the first prototype was discovered in the breakthrough paper by Francesco Calogero. They reveal a lot of interesting structures and have opened many interesting new directions admitting various extensions. In particular, the Hamiltonians of the Ruijsenaars-Schenider (RS) system \cite{RS} form a commutative subalgebra of the Ding-Iohara-Miki (DIM) algebra \cite{DI,Miki}, or equivalently, the elliptic Hall algebra \cite{K,BS,S} (which is basically the same \cite{S,Feigin}), and the simplest of the RS Hamiltonians (aka Macdonald operator \cite{Macop}) in the case of $n$ particles is
\be
H_1^{RS}=\sum_{i=1}^N\prod_{j\ne i}{tx_i-x_j\over x_i-x_j}q^{\hat D_i}\nn
\ee
where $\hat D_i:=x_i{\p\over\p x_i}$. The eigenfunctions of the RS Hamiltonians are the celebrated symmetric Macdonald polynomials \cite{Mac}.

In terms of the elliptic Hall algebra given by the generating elements $e_{\vec\gamma}$ associated with points of the integer two-dimensional lattice (i.e. $\vec\gamma$ is a vector in this lattice), the RS Hamiltonians are lying on the vertical ray $H_k^{RS}=H_k^{(0,1)}:=e_{(0,k)}$, see Fig.1 and form a commutative subalgebra. In this algebra, by virtute of two automorphisms of the elliptic Hall algebra, ${\cal O}_h$ and ${\cal O}_v$ called Miki automorphisms \cite{Miki1,Miki}, which are generators of the $SL(2,\mathbb{Z})$ group:
\begin{align}
  \mathcal{O}_h : e_{(p,r)} \rightarrow e_{(p, p-r)}\nn \\
  \mathcal{O}_v : e_{(p,r)} \rightarrow e_{(p+r, r)}\nn
\end{align}
one can generate many more commutative subalgebras, associated with rays outcoming from the origin. In the elliptic Hall algebra formulation \cite{BS}, commutativity of these rays is one of the defining axioms
\be
\left[ e_{\vec\gamma}, e_{k\vec \gamma}\right] = 0 \ \ \ \ \ \forall \vec\gamma \ {\rm and} \ k\in Z_+\nn
\ee
and the $SL(2,\mathbb{Z})$ symmetry allows one to connect these rays with each other.
The $SL(2,\mathbb{Z})$ automorphism ${\cal O}_v^{-1}$ acts in a trivial way \cite{MMP}. In particular, it immediately maps the vertical ray (i.e. the RS Hamiltonians) to ray $(-1,1)$:
\be
H_k^{(-1,1)}:=e_{[-k,k]}=q^{{1\over 2}\sum_{i=1}^n(\log_q x_i)^2}\cdot\hat H^{(0,1)}_k\cdot q^{-{1\over 2}\sum_{i=1}^n(\log_q x_i)^2}\nn
\ee
Hence, the eigenfunctions of these Hamiltonians are just the same symmetric Macdonald polynomials multiplied by the factor of $q^{{1\over 2}\sum_{i=1}^n(\log_q x_i)^2}$.

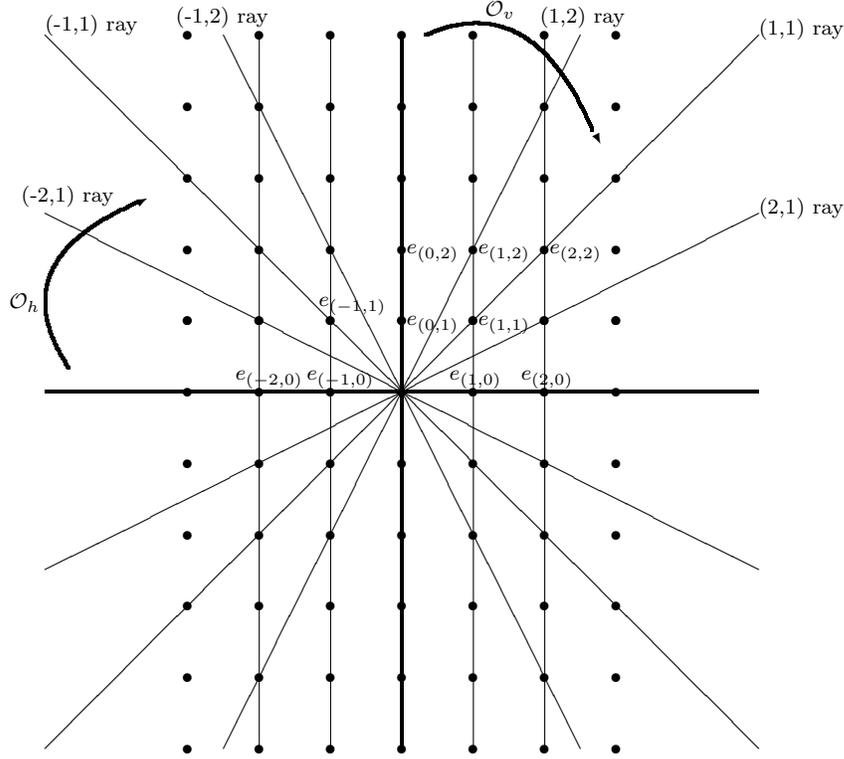
\begin{figure}[h]
\setlength{\unitlength}{.9pt}
\begin{picture}(350,340)(-285,-170)
{\footnotesize
\put(-30,-150){\line(0,1){300}}
\put(-60,-150){\line(0,1){300}}
\put(30,-150){\line(0,1){300}}
\put(60,-150){\line(0,1){300}}


\put(-32.5,147.5){\mbox{$\bullet$}}
\put(-32.5,117.5){\mbox{$\bullet$}}
\put(-32.5,87.5){\mbox{$\bullet$}}
\put(-32.5,57.5){\mbox{$\bullet$}}
\put(-32.5,27.5){\mbox{$\bullet$}}
\put(-32.5,-2.5){\mbox{$\bullet$}}
\put(-32.5,-32.5){\mbox{$\bullet$}}
\put(-32.5,-62.5){\mbox{$\bullet$}}
\put(-32.5,-92.5){\mbox{$\bullet$}}
\put(-32.5,-122.5){\mbox{$\bullet$}}
\put(-32.5,-152.5){\mbox{$\bullet$}}
\put(-32.5,27.5){\mbox{$\bullet$}}

\put(-62.5,147.5){\mbox{$\bullet$}}
\put(-62.5,117.5){\mbox{$\bullet$}}
\put(-62.5,87.5){\mbox{$\bullet$}}
\put(-62.5,57.5){\mbox{$\bullet$}}
\put(-62.5,27.5){\mbox{$\bullet$}}
\put(-62.5,-2.5){\mbox{$\bullet$}}
\put(-62.5,-32.5){\mbox{$\bullet$}}
\put(-62.5,-62.5){\mbox{$\bullet$}}
\put(-62.5,-92.5){\mbox{$\bullet$}}
\put(-62.5,-122.5){\mbox{$\bullet$}}
\put(-62.5,-152.5){\mbox{$\bullet$}}
\put(-62.5,27.5){\mbox{$\bullet$}}

\put(-92.5,147.5){\mbox{$\bullet$}}
\put(-92.5,117.5){\mbox{$\bullet$}}
\put(-92.5,87.5){\mbox{$\bullet$}}
\put(-92.5,57.5){\mbox{$\bullet$}}
\put(-92.5,27.5){\mbox{$\bullet$}}
\put(-92.5,-2.5){\mbox{$\bullet$}}
\put(-92.5,-32.5){\mbox{$\bullet$}}
\put(-92.5,-62.5){\mbox{$\bullet$}}
\put(-92.5,-92.5){\mbox{$\bullet$}}
\put(-92.5,-122.5){\mbox{$\bullet$}}
\put(-92.5,-152.5){\mbox{$\bullet$}}
\put(-92.5,27.5){\mbox{$\bullet$}}

\put(-2.5,147.5){\mbox{$\bullet$}}
\put(-2.5,117.5){\mbox{$\bullet$}}
\put(-2.5,87.5){\mbox{$\bullet$}}
\put(-2.5,57.5){\mbox{$\bullet$}}
\put(-2.5,27.5){\mbox{$\bullet$}}
\put(-2.5,-2.5){\mbox{$\bullet$}}
\put(-2.5,-32.5){\mbox{$\bullet$}}
\put(-2.5,-62.5){\mbox{$\bullet$}}
\put(-2.5,-92.5){\mbox{$\bullet$}}
\put(-2.5,-122.5){\mbox{$\bullet$}}
\put(-2.5,-152.5){\mbox{$\bullet$}}
\put(-2.5,27.5){\mbox{$\bullet$}}

\put(27.5,147.5){\mbox{$\bullet$}}
\put(27.5,117.5){\mbox{$\bullet$}}
\put(27.5,87.5){\mbox{$\bullet$}}
\put(27.5,57.5){\mbox{$\bullet$}}
\put(27.5,27.5){\mbox{$\bullet$}}
\put(27.5,-2.5){\mbox{$\bullet$}}
\put(27.5,-32.5){\mbox{$\bullet$}}
\put(27.5,-62.5){\mbox{$\bullet$}}
\put(27.5,-92.5){\mbox{$\bullet$}}
\put(27.5,-122.5){\mbox{$\bullet$}}
\put(27.5,-152.5){\mbox{$\bullet$}}
\put(27.5,27.5){\mbox{$\bullet$}}

\put(57.5,147.5){\mbox{$\bullet$}}
\put(57.5,117.5){\mbox{$\bullet$}}
\put(57.5,87.5){\mbox{$\bullet$}}
\put(57.5,57.5){\mbox{$\bullet$}}
\put(57.5,27.5){\mbox{$\bullet$}}
\put(57.5,-2.5){\mbox{$\bullet$}}
\put(57.5,-32.5){\mbox{$\bullet$}}
\put(57.5,-62.5){\mbox{$\bullet$}}
\put(57.5,-92.5){\mbox{$\bullet$}}
\put(57.5,-122.5){\mbox{$\bullet$}}
\put(57.5,-152.5){\mbox{$\bullet$}}
\put(57.5,27.5){\mbox{$\bullet$}}

\put(87.5,147.5){\mbox{$\bullet$}}
\put(87.5,117.5){\mbox{$\bullet$}}
\put(87.5,87.5){\mbox{$\bullet$}}
\put(87.5,57.5){\mbox{$\bullet$}}
\put(87.5,27.5){\mbox{$\bullet$}}
\put(87.5,-2.5){\mbox{$\bullet$}}
\put(87.5,-32.5){\mbox{$\bullet$}}
\put(87.5,-62.5){\mbox{$\bullet$}}
\put(87.5,-92.5){\mbox{$\bullet$}}
\put(87.5,-122.5){\mbox{$\bullet$}}
\put(87.5,-152.5){\mbox{$\bullet$}}
\put(87.5,27.5){\mbox{$\bullet$}}

\put(150,-75){\line(-2,1){300}}
\put(-150,-75){\line(2,1){300}}
\put(75,-150){\line(-1,2){150}}
\put(-75,-150){\line(1,2){150}}

\put(150,150){\mbox{(1,1) ray}}
\put(150,75){\mbox{(2,1) ray}}
\put(-150,152){\mbox{(-1,1) ray}}
\put(-160,80){\mbox{(-2,1) ray}}
\put(58,155){\mbox{(1,2) ray}}
\put(-95,155){\mbox{(-1,2) ray}}

\put(-40,5){\mbox{$e_{(-1,0)}$}}
\put(-70,5){\mbox{$e_{(-2,0)}$}}
\put(20,5){\mbox{$e_{(1,0)}$}}
\put(50,5){\mbox{$e_{(2,0)}$}}

\put(-35,36){\mbox{$e_{(-1,1)}$}}
\put(32,28){\mbox{$e_{(1,1)}$}}
\put(2,28){\mbox{$e_{(0,1)}$}}
\put(2,58){\mbox{$e_{(0,2)}$}}
\put(32,58){\mbox{$e_{(1,2)}$}}
\put(62,58){\mbox{$e_{(2,2)}$}}

\linethickness{1.5pt}
\put(150,-150){\line(-1,1){300}}
\put(-150,-150){\line(1,1){300}}
\put(0,-150){\line(0,1){300}}
\put(-150,0){\line(1,0){300}}

\qbezier(10,150)(55,170)(80,110)
\put(80.5,110){\vector(1,-2){3}}
\put(35,158){\mbox{${\cal O}_v$}}

\qbezier(-140,10)(-170,55)(-110,80)
\put(-110,80){\vector(2,1){3}}
\put(-165,35){\mbox{${\cal O}_h$}}

}
\end{picture}
\caption{\footnotesize $2d$ integer lattice of generators of the elliptic Hall/DIM algebra. Each ray (p,r) gives rise to a commutative subalgebra, and each pair of rays (p,r) and (-p,-r) form a Heisenberg subalgebra.}
\label{Dimfig}
\end{figure}

However, action of the second automorphism ${\cal O}_h$ is far less trivial, and the commutative Hamiltonians associated with rays $(-1,a)$, which can be obtained by action of this automorphism are not that simple. One of the ways to generate them explicitly is described in \cite{MMPBA2}. The simplest Hamiltonian at $a=2$ is \cite{CF,MMPBA2}
\be
\hat H^{(-1,2)}_1:=e_{[-1,2]}&= & \ \sum_{i=1}^N \frac{1}{q^{1\over 2}x_i}
    \prod_{j \neq i} \frac{(t x_i - x_j)}{(x_i - x_j)}\frac{(qt x_i - x_j)}{(qx_i - x_j)}
    \ttop{i}\ttop{i}
    \nn \\
    & + &{q^{1\over 2}(t - q)(t - 1)\over q-1}
    \sum_{i \neq j} \prod_{k \neq i,j}
    \frac{(t x_i - x_k)(t x_j - x_k)}{(x_i - x_k)(x_j - x_k)}
    \frac{1}{(q x_i - x_j)} \ttop{i} \ttop{j}\nn
\ee
As we already noted, eigenfunctions of the Hamiltonians $\hat H^{(-1,1)}_k$ associated with ray (-1,1) are the same symmetric Macdonald polynomials multiplied by the factor of $q^{{1\over 2}\sum_{i=1}^n(\log_q x_i)^2}$.
Eigenfunctions of the Hamiltonians $\hat H^{(-1,a)}_k$ are \cite{CF,MMPBA1} the so called twisted Baker-Akhiezer functions introduced by O. Chalykh \cite{Cha,CE} multiplied by the factor $q^{{1\over 2a}\sum_{i=1}^n(\log_q x_i)^2}$.
The Baker-Akhiezer function is a (quasi)polynomial. Hence, in order to deal with polynomials, we will use the ``rotated'' Hamiltonians defined as
\be\label{rotH}
\hat {\cal H}^{(a)}_k=q^{-{1\over 2a}\sum_{i=1}^n(\log_q x_i)^2}\cdot\underbrace{\hat H^{(-1,a)}_k}_{e_{[-k,ak]}}
\cdot q^{{1\over 2a}\sum_{i=1}^n(\log_q x_i)^2}
\ee

The next important step is to note that the Hamiltonians of all these commutative subalgebras associated with rays can be realized in terms of power sums of expressions made of Cherednik operators and {\bf acting on the space of symmetric functions}\footnote{At the algebraic level, it expresses the relation \cite{DIMDAHA} between the DIM (Elliptic Hall) algebra and spherical DAHA \cite{Ch}.} \cite{MMP}\footnote{Note that all formulas in \cite{MMP} differ from those in the present paper by the rotation as in (\ref{rotH}).}. We call these expressions twisted Cherednik operators, and they give rise to new systems of commuting Hamiltonians and, hence, we arrive at new many-body integrable systems. They are of our main interest here. Note that these commutative twisted Cherednik operators are generally acting on spaces of {\bf non-symmetric} functions.

In particular, these Hamiltonians associated with the simplest ray $(-1,1)$ are just the Cherednik Hamiltonians $C_i$: ${\cal H}_k^{(-1,1)}=\sum_{i=1}^n C_i^k$. In the case of $n=2$, they are
\be
C_1={(tx_1-x_2)\over t(x_1-x_2)}q^{\hat D_1}+{(1-t)x_2\over t(x_1-x_2)}\sigma_{1,2}q^{\hat D_1}\nn\\
C_2=q^{\hat D_2}{(x_1-tx_2)\over t(x_1-x_2)}-q^{\hat D_2}{(1-t)x_2\over t(x_1-x_2)}\sigma_{1,2}\nn
\ee
where $\sigma_{1,2}$ permutes $x_1$ and $x_2$.
Eigenfunctions of these Hamiltonians are the non-symmetric Macdonald polynomials \cite{Opd95,Mac96,Che95}.

Similarly, the Hamiltonians associated with rays $(-1,a)$ are the twisted Cherednik Hamiltonians ${\mathfrak{C}}_i^{(a)}$: ${\cal H}_k^{(-1,a)}=\sum_{i=1}^n \left({\mathfrak{C}}_i^{(a)}\right)^k$. In the case of $n=2$, $a=2$, they are (see details in sec.2.2)
\be
{\mathfrak{C}}_1^{(2)}&=&{q^{1\over 2}\over t^2}{(tx_1-x_2)(qtx_1-x_2)\over(x_1-x_2)(qx_1-x_2)}q^{2\hat D_1}+
{q^{1\over 2}\over t^2}{(1-t)x_2(x_1-qtx_2)\over(x_1-x_2)(x_1-qx_2)}q^{2\hat D_2}\sigma_{1,2}\nn\\
&+&{q\over t^2}{(1-t)(tx_1-x_2)x_1^{1\over 2}x_2^{1\over 2}\over(x_1-x_2)(qx_1-x_2)}q^{\hat D_1+\hat D_2}\sigma_{1,2}-
{q\over t^2}{(1-t)^2x_1^{1\over 2}x_2^{3\over 2}\over(x_1-x_2)(x_1-qx_2)}q^{\hat D_1+\hat D_2}
\nn\\
{\mathfrak{C}}_2^{(2)}&=&q^{\hat D_2}{q^{1\over 2}\over t^2}{(x_1-tx_2)(x_1-qtx_2)\over(x_1-x_2)(x_1-qx_2)}q^{\hat D_2}-
q^{\hat D_2}{q^{3\over 2}\over t^2}{(1-t)(x_1-tx_2)x_2\over(x_1-x_2)(x_1-qx_2)}q^{\hat D_2}\sigma_{1,2}-\nn\\
&-&{q\over t^2}{(1-t)(tx_1-x_2)x_1^{1\over 2}x_2^{1\over 2}\over(x_1-x_2)(x_1-qx_2)}q^{\hat D_1+\hat D_2}\sigma_{1,2}-
{q\over t^2}{(1-t)^2x_1^{3\over 2}x_2^{1\over 2}\over(x_1-x_2)(x_1-qx_2)}q^{\hat D_1+\hat D_2}\nn
\ee
The eigenfunctions of these twisted Cherednik Hamiltonians are then naturally called non-symmetric twisted Macdonald polynomials \cite{MMP1,MMP2,MMP3}. Quite non-trivially, the sums of these eigenfunctions with proper coefficients produce symmetric functions, which are eigenfunctions of the original DIM Hamiltonians $\hat {\cal H}^{(a)}_k$ \cite{MMP3}, which is in complete analogy with the standard fact that the non-symmetric Macdonald polynomials being summed to the symmetric ones are eigenfunctions of the RS Hamiltonians.

In this paper, we study the twisted Cherednik integrable system, and our goal is to describe the eigenfunctions of the twisted Cherednik
Hamiltonians. We start in section 2 with description of the Cherednik and twisted Cherednik Hamiltonians, and describe a set of algebraic relations giving them. Then, in sections 3 and 4, we obtain eigenfunctions of the Cherednik and twisted Cherednik Hamiltonians correspondingly. In section 5, we describe essential properties of the twisted eigenfunctions, and, in section 6, an algorithmic procedure, which allows one to generate these eigenfunctions effectively. Section 7 contains a summary and some concluding remarks. We attach to this submission a MAPLE file that allows one to generate the twisted and non-twisted Macdonald polynomials, both non-symmetric and symmetric, its short description can be found in the Appendix.

\paragraph{Notation.} Throughout the paper, we use the notation $\alpha$ for the weak composition (of length $n$) of an integer $|\alpha|:=\sum_{i=1}^n\alpha_i$ (we admit some $\alpha_i=0$ may stand at the end of the weak composition), the notation $\alpha^+$ for the corresponding partition (Young diagram), i.e. the \wc when $\alpha_i$ are ordered: $\alpha_1\ge\alpha_2\ge\ldots\ge\alpha_n\ge 0$, and the notation $\alpha^-$ for the inversely ordered case $0\le\alpha_1\le\alpha_2\le\ldots\le\alpha_n$.

We also regularly use the following quantities: the eigenvalues of the Cherednik Hamiltonians
\be
\Lambda^{(i)}_\alpha=q^{\alpha_i}t^{-\zeta(\alpha)_i}\nn
\ee
where $\zeta(\alpha)_i=\#\{k<i|\alpha_k\ge\alpha_i\}+\#\{k>i|\alpha_k>\alpha_i\}$, and their ratios
\be
r_{\alpha,i}:={\Lambda_\alpha^{(i+1)}\over\Lambda_\alpha^{(i)}}=q^{\alpha_{i+1}-\alpha_i}t^{\zeta(\alpha)_i-\zeta(\alpha)_{i+1}}\nn
\ee
Note that $\zeta(\alpha^+)_i=i-1$.

Throughout the paper, the term ``polynomial'' implies a polynomial in $x_i^{1\over a}$'s.

We also use the notation
\be
\{x\}:=1-x
\ee

\section{Twisted Cherednik Hamiltonians}

We start with description of operators and their commutation relations used in constructing the Cherednik system.

\subsection{Basic operators \label{bops}}

Cherednik operators $C_k$ and Demazure-Lustig operators $T_i$ are defined \cite{Ch,NSCh,BF}
\be
R_{ij}:&=&1+{(1-t^{-1})x_j\over x_i-x_j}(1-\sigma_{i,j})\\
R_{ij}^{-1}&=&1+{(1-t)x_j\over x_i-x_j}(1-\sigma_{i,j})\nn\\
T_i:&=&R_{i,i+1}\sigma_{i,i+1}=\sigma_{i,i+1}+{(t^{-1}-1)x_{i+1}\over x_i-x_{i+1}}(1-\sigma_{i,i+1})=
1+{x_i-t^{-1}x_{i+1}\over x_i-x_{i+1}}(\sigma_{i,i+1}-1),\ \ \ \ \ i=1,\ldots,n-1\nn\\
T_i^{-1}&=&\sigma_{i,i+1}R_{i,i+1}^{-1}\nn\\
C_i: &=& t^{1 - i} \left(\prod_{j = i + 1}^n R_{i,j}\right)
    q^{\hat D_i}
    \left(\prod_{j = 1}^{i - 1} R^{-1}_{j,i}\right)
    \nn\\
B:&=&T_{n-1}\ldots T_2T_1 x_1\nn
\label{ordefs}
\ee
where $\hat D_i:=x_i{\p\over\p x_i}$ and $\sigma_{i,j}$ permutes $x_i$ and $x_j$. The products in $C_i$ are obtained so that the smaller index stands to the left.

One can also introduce the operation $\pi$:
\be
\pi F(x_1,x_2,\ldots,x_n):=F(qx_n,x_1,\ldots,x_{n-1})
\ee
so that
\be\label{Cpi}
C_i=t^{1-i}T_iT_{i+1}\ldots T_{n-1}\pi T_1^{-1}T_2^{-1}\ldots T_{i-1}^{-1}
\ee

These quantities satisfy a set of relations:
\begin{itemize}
\item At $i=1,\ldots,n-2$ (Hecke algebra):
\be
(T_i-1)(T_i+t^{-1})&=&0\nn\\
\phantom{.}[T_i,T_j]&=&0,\ \ \ \ \ \ \ |i-j|\ge 2\\
T_iT_{i+1}T_i&=&T_{i+1}T_iT_{i+1}
\ee
\item At $i=1,\ldots,n-1$:
\be\label{TC}
tT_iC_{i+1}T_i&=&C_i\nn\\
\phantom{.}[T_i,C_j]&=&0\ \ \ \ \ \ i\ne j,j-1
\ee
\be\label{Tx}
tT_ix_iT_i&=&x_{i+1}\nn\\
\phantom{.}[T_i,x_j]&=&0\ \ \ \ \ \ i\ne j,j+1
\ee
\be\label{CB}
C_iB=B C_{i+1}
\ee
\item At $i=1,\ldots,n-2$:
\be
\pi^{-1}T_i\pi=T_{i+1}
\ee
\item At $i,j=1,\ldots,n$:
\be
\phantom{.}[C_i,C_j]=0
\ee
\end{itemize}

\subsection{Twisted Hamiltonians}

We introduce $a$-twisted Hamiltonians ${\mathfrak{C}}_i^{(a)}$, their power sums being associated with the ray $(-1,a)$ DIM Hamiltonians, which reduce to the standard Cherednik ones at $a=1$, and have zero grading\footnote{We ``rotate'' the $a$-twisted Hamiltonians as compared with \cite[Sec.6]{MMP} so that they have zero grading and may admit polynomial solutions. This is the origin of peculiar $x_i^{1-a\over a}$-factor in front of $q^{\hat D_i}$, in contrast with the naive $x^{-1}_i$.} in $x_i$'s at any $a$:
\be
{\cal C}_i^{(a)}&=&
t^{1 - i} \left(\prod_{j = i + 1}^n R_{i,j}\right)
    {1\over x_i^{a-1\over a}}q^{\hat D_i}
    \left(\prod_{j = 1}^{i - 1} R^{-1}_{j,i}\right)\nn\\
   {\mathfrak{C}}_i^{(a)}:&=&{1\over x_i}\Big(x_i{\cal C}_i^{(a)}\Big)^a
\ee
The corresponding twisted ``intertwining" operator is
\be
\mathfrak{B}^{(a)}=\left(T_{n-1}\ldots T_2T_1\pi^{-1}x_n^{-{1\over a}}\right)^{a-1}\cdot B=q\left(T_{n-1}\ldots T_2T_1x_1^{-{1\over a}}\pi^{-1}\right)^a\pi x_1^{a+1\over a}
\ee

They celebrate the following properties:
\begin{itemize}
\item At $i=1,\ldots,n-1$:
\be\label{TC2}
tT_i{\mathfrak{C}}_{i+1}^{(a)}T_i&=&{\mathfrak{C}}_i^{(a)}\nn\\
\phantom{.}[T_i,{\mathfrak{C}}_j^{(a)}]&=&0\ \ \ \ \ \ i\ne j,j-1
\ee
\be\label{BCa}
\mathfrak{B}^{(a)}{\mathfrak{C}}_{i+1}^{(a)}={\mathfrak{C}}_i^{(a)}\mathfrak{B}^{(a)}
\ee
\item At $i,j=1,\ldots,n$:
\be
\phantom{.}[{\mathfrak{C}}_i^{(a)},{\mathfrak{C}}_j^{(a)}]=0
\ee
\end{itemize}

\section{Eigenfunctions of the Cherednik Hamiltonians}

\subsection{Non-symmetric Macdonald polynomials}

The common polynomial eigenfunctions of $C_i$'s are enumerated by \wcs (i.e. compositions admitting zero numbers) and are called non-symmetric Macdonald polynomials $E_{w\lambda}$, where $\lambda$ is the Young diagram, i.e. $\lambda_1\ge\lambda_2\ge\ldots\ge\lambda_n\ge 0$, and $w$ is a permutation:
\be
C_i\cdot E_{w\lambda}=\Lambda^{(i)}_{w\lambda}\cdot E_{w\lambda}\nn
\ee
The eigenvalues are
\be\label{ev}
\Lambda^{(i)}_\lambda=q^{\lambda_i}t^{1-i}
\ee
and
\be\label{evw}
\Lambda^{(i)}_{w\lambda}=\Lambda^{w([1,n])_i}_\lambda
\ee
where we denoted through $w([1,n])_i$ the $i$-th element of the sequence $w([1,n])$.

If one denotes $\alpha:=w\lambda$, it can be also written in the form
\be\label{evalpha}
\Lambda^{(i)}_\alpha=q^{\alpha_i}t^{-\zeta(\alpha)_i}
\ee
where $\zeta(\alpha)_i:=\#\{k<i|\alpha_k\ge\alpha_i\}+\#\{k>i|\alpha_k>\alpha_i\}$.

The non-symmetric Macdonald polynomials are graded with the grading $|\alpha|:=\sum_{i=1}^n\alpha_i$.

\subsection{Triangular structure}

The non-symmetric Macdonald polynomials are
\be\label{nsM}
E_{\alpha}=x^\alpha+\sum_{\beta<\alpha}C_{\alpha\beta}x^\mu
\ee
where $\alpha$ is a \wc with $n$ parts (unordered, and some of the parts may be zero). If there are two \wcs, $\alpha$ and $\beta$, $\alpha>\beta$ if $\alpha^+>\beta^+$ (e.g., in accordance with the lexicographic order), and if the ordered partitions coincide, one compares the minimal length of permutations of the symmetric group ${\cal S}_n$ that allow one to make an ordered partition. The less is the length, the larger is \wc. In other words, the largest one is the Young diagram. This is called {\it Bruhat order} \cite{HHL}.

One of the ways to unambiguously restore the coefficients $C_{\alpha\beta}$ in (\ref{nsM}) is to use an orthogonality condition with respect to the Cherednik scalar product:
\be
\Big<f,g\Big>=\prod_{i=1}^n\oint{dx_i\over x_i}f(x_i;q,t)g(x_i^{-1};q^{-1},t^{-1})\prod_{i>j}{(x_i/x_j;q)_\infty(qx_j/x_i;q)_\infty
\over (tx_i/x_j;q)_\infty(tqx_j/x_i;q)_\infty}\nn
\ee

\subsection{Properties of non-symmetric Macdonald polynomials}

The operator $B$ acts on the non-symmetric Macdonald polynomials in the following way \cite{BF}:
\be\label{B}
B\cdot E_{[\alpha_1,\ldots,\alpha_n]}(x_1,\ldots,x_n)=t^{-\#\{\alpha_i\le\alpha_1\}}
E_{[\alpha_2,\ldots,\alpha_{n},\alpha_1+1]}(x_1,\ldots,x_n)
\ee
One can immediately see this from (\ref{CB}), (\ref{evalpha}) and the fact that action of $B$ raises the grading up by unity, and the coefficient is restored from the analysis of the leading term in the triangular expansion (\ref{nsM}).

Acting on the both parts of (\ref{B}) by $C_n$ in the form (\ref{Cpi}) and using (\ref{evalpha}), one comes to the symmetry property of the non-symmetric Macdonald polynomials (Knop–Sahi recurrence) \cite{KS,HHL}
\be\label{Bs}
E_{[\alpha_2,\ldots,\alpha_n,\alpha_1+1]}(x_1,x_2,\ldots,x_n)=
q^{-\alpha_1}x_nE_{[\alpha_1\alpha_2,\ldots,\alpha_n]}(qx_{n},x_1,x_2,\ldots,x_{n-1})
\ee

Other essential properties of the non-symmetric Macdonald polynomials are valid also for the twisted polynomials (this is because commutation relations of ${\mathfrak{C}}_i^{(a)}$ and $C_i$ with $T_i$ are the same, see (\ref{TC}) and (\ref{TC2})), we discuss them in sec.\ref{PM} for arbitrary $a$, and do not repeat them here. They include:
\begin{itemize}
\item The stability property, i.e. reduction under $x_n=0$, see (\ref{stab}).
\item Their behaviour under the action of operators $T_i$'s, see (\ref{perm}).
\item Construction of the symmetric polynomials, see (\ref{nss}).
\end{itemize}

\section{Eigenfunctions of the twisted Cherednik Hamiltonians}

The polynomial solutions of the twisted Cherednik Hamiltonians ${\mathfrak{C}}_i^{(a)}$ are available at $t=q^{-m}$ with $m$ being natural. Hence, hereafter we always imply $t$ of this form.

\subsection{Ground state solution}

The simplest of the eigenfunctions is the ground state, which is a symmetric function, as usual for the ground state eigenfunctions. Since symmetric eigenfunctions of the twisted Cherednik Hamiltonians ${\mathfrak{C}}_i^{(a)}$ are simultaneously eigenfunctions of the DIM Hamiltonians ${\cal H}_k^{(a)}$ \cite{MMP}\footnote{Our definition of the twisted Cherednik Hamiltonians ${\mathfrak{C}}_i^{(a)}$ associates them with the DIM Hamiltonians ``rotated'' by the factor $q^{{1\over 2a}\sum_{i=1}^n(\log_q x_i)^2}$, see \cite[Secs.2.2-2.3, and Eq.(12)]{MMP3}.
}, we start with constructing the proper solution to these latter. It is available at $t=q^{-m}$ in the polynomial form and is called twisted Baker-Akhiezer function \cite{CE,CF,MMPBA1}.

The twisted Baker-Akhiezer function, which is a function of $2N$ complex parameters $x_i$ and $y_i$, $i=1,\ldots,N$, is defined as a sum
\be
\mathfrak{B}_m^{(a)}(\vec x,\vec y)=\left(\prod_{i=1}^Nx_i^{{\log_q y_i\over a}+m\rho_i}\right)\cdot \sum_{k_{ij}=0}^{ma}\prod_{i<j}\left({x_j\over x_i}\right)^{k_{ij}\over a}b^{(a)}_{m,\vec y,k}\nn
\ee
with the property
\be
\mathfrak{B}_m^{(a)}(x_kq^j,\vec y)=\varepsilon^j\mathfrak{B}_m^{(a)}(x_lq^j,\vec y)\ \ \ \ \  \forall k,l\ \ \hbox{and}\ \ 1\le j\le m\ \ \ \ \ \hbox{at}\ \ \varepsilon x_k^{1\over a}= x_l^{1\over a}\nn
\ee
for any $\varepsilon$ such that $\varepsilon^a=1$. Here $\vec\rho$ is the Weyl vector, i.e. $\rho_i={1\over 2}(N-2i+1)$. This $a$-twisted Baker-Akhiezer function is unique up to normalization, and, upon a proper normalization, is symmetric with respect to the permutation of $\vec x$ and $\vec y$.
Note that the generic Baker-Akhiezer function is a quasipolynomial because of the common monomial factor. However, we will be interested in this paper only in the polynomial Baker-Akhiezer functions: they are polynomials in $x_i^{1\over a}$'s.

The ground state eigenfunction of the twisted Cherednik Hamiltonians ${\mathfrak{C}}_i^{(a)}$ is $\mathfrak{B}_m^{(a)}(\vec x,\vec y)$ at peculiar values of $\vec y$:
\be\label{omegaBA}
\boxed{\begin{array}{c}
\Omega_m^{(a)}(\vec x)=\mathfrak{B}_m^{(a)}(\vec x,\vec y^\ast)\cr\cr
y_i^\ast=q^{(i-1)m+m{(a-2)(N-1)\over 2}}
\end{array}
}
\ee
This is a polynomial of variables $x_i^{1\over a}$ of degree $d_0=\frac{N(N-1)}{2}\cdot (a-1)\cdot m$.

At $a=1$, all $\Omega_m^{(1)}(\vec x)=1$. At $a=2$, $N=2$, it is given by a simple product:
\be\label{omega22}
\Omega_m^{(a)}(x_1,x_2)&=&\prod_{j=0}^{m-1}\Big(x_1^{1\over 2}+q^{j-\frac{m-1}{2}}x_2^{1\over 2} \Big)
\ee
At $N=2$ and arbitrary $a$, the expression is more involved \cite[Eq.(94)]{MMP1}, and generally the formulas for $\Omega_m^{(a)}(\vec x)$ become very tedious \cite{MMP2}.

\subsection{Twisted non-symmetric Macdonald polynomials as eigenfunctions of the Cherednik operators}

The excitation eigenfunctions are all described basing on this ground state. Similarly to the non-symmetric Macdonald polynomials, they are enumerated by the \wcs $\alpha$, hence, the name twisted non-symmetric Macdonald polynomials. They solve the equation
\be\label{ee}
{\mathfrak{C}}_i^{(a)}\cdot E^{(a)}_{\alpha}=\Lambda^{(i,a)}_{\alpha}\cdot E^{(a)}_{\alpha}
\ee
where the eigenvalues are just rescaled non-twisted eigenvalues:
\be\label{eva}
\Lambda^{(i,a)}_\alpha=q^{a-1\over 2}t^{2(1-a)}\Lambda^{(i)}_\alpha
\ee
It is convenient to introduce polynomials
\be\label{Xi}
\Xi^{(a)}_{\alpha}:=\left(\prod_{i=1}x_i^{\alpha_i\over a}q^{\alpha_i(\alpha_i-1)\over 2a}\right)\ \Omega^{(a)}(q,t;\{q^{\alpha_i}x_i\})=
\prod_{i=1}\left(x_i^{1\over a}q^{\hat D_i}\right)^{\alpha_i}\ \Omega^{(a)}(q,t;\{x_i\})
\ee
then any polynomial solution to (\ref{ee}) can be presented in the form of {\bf a linear combination of such $\Xi^{(a)}_{\alpha}$ with coefficients that are rational functions independent of the twist $a$} \cite{MMP1,MMP2,MMP3}:
\be\label{gE}
\boxed{
E^{(a)}_{\alpha}(\vec x)=\sum_{\beta\le\alpha} F_{\alpha,\beta}(\vec x)\cdot \Xi^{(a)}_{\beta}
}
\ee
where $F_{\alpha,\beta}$ does not depend on $a$ at all \cite{MMP1}. Thus, the twisted non-symmetric Macdonald polynomial have also a triangular structure similar to the non-twisted ones (\ref{nsM}).

\subsection{Examples}

In order to get a flavour of what the solutions look like, we list here first few examples at $n=3$. Hereafter, we fix the normalization of functions $F_{\alpha,\beta}(\vec x)$ in the following way. Since $F_{\alpha,\beta}(\vec x)$'s do not depend on $a$, it is sufficient to fix it at $a=1$ in (\ref{gE}). Thus, we put $\Omega^{(a)}(\vec x)=1$ so that $E_\alpha^{(1)}$ can be presented in the triangular form (\ref{nsM}), and then we require the leading term coefficient in (\ref{nsM}) to be unity.

\paragraph{Young diagram [1].}

\be\label{n3}
E_{[0,0,1]}^{(a)}&=&\Xi^{(a)}_{[0,0,1]}\nn\\
tE_{[0,1,0]}^{(a)}&=&{\Big\{{tx_2\over x_3}\Big\}\over \Big\{{x_2\over x_3}\Big\}}
\Xi^{(a)}_{[0,1,0]}+
{(1-t)\over 1-qt^2}{\Big\{{qt^2x_3\over x_2}\Big\}\over \Big\{{x_3\over x_2}\Big\}}\Xi^{(a)}_{[0,0,1]}\nn
\\
t^2E_{[1,0,0]}^{(a)}&=&{\Big\{{tx_1\over x_2}\Big\}\Big\{{tx_1\over x_3}\Big\}\over \Big\{{x_1\over x_2}\Big\}\Big\{{x_1\over x_3}\Big\}}\Xi^{(a)}_{[1,0,0]}+
{(1-t)\over (1-qt)}{\Big\{{tx_2\over x_3}\Big\}\Big\{{qtx_2\over x_1}\Big\}\over \Big\{{x_2\over x_3}\Big\}\Big\{{x_2\over x_1}\Big\}}\Xi^{(a)}_{[0,1,0]}+{(1-t)\over (1-qt)}{\Big\{{tx_3\over x_2}\Big\}\Big\{{qtx_3\over x_1}\Big\}\over \Big\{{x_3\over x_1}\Big\}\Big\{{x_3\over x_2}\Big\}}\Xi^{(a)}_{[0,0,1]}
\ee

\paragraph{Young diagram [1,1].}

\be
E_{[0,1,1]}^{(a)}&=&\Xi^{(a)}_{[0,1,1]}\\
tE_{[1,0,1]}^{(a)}&=&{\Big\{{tx_1\over x_2}\Big\}\over \Big\{{x_1\over x_2}\Big\}}\Xi^{(a)}_{[1,0,1]}
+{(1-t)\over (1-qt^2)}{\Big\{{qt^2x_2\over x_1}\Big\}\over \Big\{{x_2\over x_1}\Big\}}\Xi^{(a)}_{[0,1,1]}
\nn\\
t^2E_{[1,1,0]}^{(a)}&=&{\Big\{{tx_1\over x_3}\Big\}\over \Big\{{x_1\over x_3}\Big\}}
{\Big\{{tx_2\over x_3}\Big\}\over \Big\{{x_2\over x_3}\Big\}}\ \Xi^{(a)}_{[1,1,0]}+
{(1-t)\over (1-qt)}\left(
{\Big\{{tx_1\over x_2}\Big\}\over \Big\{{x_1\over x_2}\Big\}}{\Big\{{qtx_3\over x_2}\Big\}\over \Big\{{x_3\over x_2}\Big\}}\
\Xi^{(a)}_{[1,0,1]}+{\Big\{{tx_2\over x_1}\Big\}\over \Big\{{x_2\over x_1}\Big\}}
{\Big\{{qtx_3\over x_1}\Big\}\over \Big\{{x_3\over x_1}\Big\}}\ \Xi^{(a)}_{[0,1,1]}\right)\nn
\ee

\paragraph{Young diagram [2].}

\be\label{l2}
qt^2E_{[0,0,2]}^{(a)}&=&{\Big\{{qtx_3\over x_2}\Big\}\over \Big\{{qx_3\over x_2}\Big\}}{\Big\{{qtx_3\over x_1}\Big\}\over\Big\{{qx_3\over x_1}\Big\}}\Xi^{(a)}_{[0,0,2]}+{(1-t)\over (1-qt)}
\left({\Big\{{tx_1\over x_3}\Big\}\over \Big\{{x_1\over qx_3}\Big\}}{\Big\{{tx_1\over x_2}\Big\}\over\Big\{{x_1\over x_2}\Big\}}\Xi^{(a)}_{[1,0,1]}+{\Big\{{tx_2\over x_3}\Big\}\over \Big\{{x_2\over qx_3}\Big\}}{\Big\{{tx_2\over x_1}\Big\}\over\Big\{{x_2\over x_1}\Big\}}\Xi^{(a)}_{[0,1,1]}\right)\nn\\
qt^3E_{[0,2,0]}^{(a)}&=&{\Big\{{qtx_2\over x_3}\Big\}\over \Big\{{qx_2\over x_3}\Big\}}{\Big\{{qtx_2\over x_1}\Big\}\over\Big\{{qx_2\over x_1}\Big\}}{\Big\{{tx_2\over x_3}\Big\}\over\Big\{{x_2\over x_3}\Big\}}\Xi^{(a)}_{[0,2,0]}+{(1-t)\over (1-q^2t^2)}
{\Big\{{q^2t^2x_3\over x_2}\Big\}\over \Big\{{qx_3\over x_2}\Big\}}{\Big\{{qtx_3\over x_2}\Big\}\over\Big\{{x_3\over x_2}\Big\}}{\Big\{{qtx_3\over x_1}\Big\}\over\Big\{{qx_3\over x_1}\Big\}}\Xi^{(a)}_{[0,0,2]}+\nn\\
&+&{(1-t)\over (1-qt)}\left({\Big\{{tx_2\over x_3}\Big\}\over \Big\{{x_2\over x_3}\Big\}}{\Big\{{tx_3\over x_2}\Big\}\over \Big\{{x_3\over qx_2}\Big\}}{\Big\{{tx_3\over x_1}\Big\}\over \Big\{{x_3\over x_1}\Big\}}+{(1-t)\over (1-q^2t^2)}
{\Big\{{q^2t^2x_3\over x_2}\Big\}\over \Big\{{x_3\over x_2}\Big\}}{\Big\{{tx_2\over x_3}\Big\}\over \Big\{{x_2\over qx_3}\Big\}}
{\Big\{{tx_2\over x_1}\Big\}\over \Big\{{x_2\over x_1}\Big\}}\right)\Xi^{(a)}_{[0,1,1]}+\nn\\
&+&{(1-t)^2\over (1-qt)(1-q^2t^2)}{\Big\{{q^2t^2x_3\over x_2}\Big\}\over \Big\{{x_3\over x_2}\Big\}}{\Big\{{tx_1\over x_3}\Big\}\over\Big\{{x_1\over qx_3}\Big\}}{\Big\{{tx_1\over x_2}\Big\}\over \Big\{{x_1\over x_2}\Big\}}\Xi^{(a)}_{[1,0,1]}+
{(1-t)\over(1-qt)}
{\Big\{{tx_2\over x_3}\Big\}\over \Big\{{x_2\over x_3}\Big\}}{\Big\{{tx_1\over x_3}\Big\}\over\Big\{{x_1\over x_3}\Big\}}{\Big\{{tx_1\over x_2}\Big\}\over \Big\{{x_1\over qx_2}\Big\}}\Xi^{(a)}_{[1,1,0]}\nn\\
qt^4E_{[2,0,0]}^{(a)}&=&{\Big\{{qtx_1\over x_3}\Big\}\over \Big\{{qx_1\over x_3}\Big\}}{\Big\{{qtx_1\over x_2}\Big\}\over\Big\{{qx_1\over x_2}\Big\}}{\Big\{{tx_1\over x_2}\Big\}\over \Big\{{x_1\over x_2}\Big\}}{\Big\{{tx_1\over x_3}\Big\}\over \Big\{{x_1\over x_3}\Big\}}\Xi^{(a)}_{[2,0,0]}+{(1-t)\over(1-q^2t)}\left(
{\Big\{{tx_2\over x_3}\Big\}\over \Big\{{x_2\over x_3}\Big\}}{\Big\{{qtx_2\over x_3}\Big\}\over\Big\{{qx_2\over x_3}\Big\}}{\Big\{{qtx_2\over x_1}\Big\}\over \Big\{{qx_2\over x_1}\Big\}}{\Big\{{q^2tx_2\over x_1}\Big\}\over \Big\{{x_2\over x_1}\Big\}}\Xi^{(a)}_{[0,2,0]}+\right.\nn\\
&+&\left.{\Big\{{tx_3\over x_2}\Big\}\over \Big\{{x_3\over x_2}\Big\}}{\Big\{{qtx_3\over x_2}\Big\}\over\Big\{{qx_3\over x_2}\Big\}}{\Big\{{qtx_3\over x_1}\Big\}\over \Big\{{qx_3\over x_1}\Big\}}{\Big\{{q^2tx_3\over x_1}\Big\}\over \Big\{{x_3\over x_1}\Big\}}\Xi^{(a)}_{[0,0,2]}\right)+{q(1+q)(1-t)^2\over(1-qt)(1-q^2t)}
{\Big\{{tx_3\over x_2}\Big\}\over \Big\{{qx_3\over x_2}\Big\}}{\Big\{{tx_2\over x_3}\Big\}\over\Big\{{qx_2\over x_3}\Big\}}{\Big\{{qtx_3\over x_1}\Big\}\over \Big\{{x_3\over x_1}\Big\}}{\Big\{{qtx_2\over x_1}\Big\}\over \Big\{{x_2\over x_1}\Big\}}\Xi^{(a)}_{[0,1,1]}+\nn\\
&+&{q(1+q)(1-t)\over (1-q^2t)}
\left({\Big\{{tx_3\over x_2}\Big\}\over \Big\{{x_3\over x_2}\Big\}}{\Big\{{tx_1\over x_3}\Big\}\over\Big\{{qx_1\over x_3}\Big\}}{\Big\{{qtx_3\over x_1}\Big\}\over \Big\{{qx_3\over x_1}\Big\}}{\Big\{{tx_1\over x_2}\Big\}\over \Big\{{x_1\over x_2}\Big\}}\Xi^{(a)}_{[1,0,1]}+
{\Big\{{tx_2\over x_3}\Big\}\over \Big\{{x_2\over x_3}\Big\}}{\Big\{{tx_1\over x_2}\Big\}\over\Big\{{qx_1\over x_2}\Big\}}{\Big\{{qtx_2\over x_1}\Big\}\over \Big\{{qx_2\over x_1}\Big\}}{\Big\{{tx_1\over x_3}\Big\}\over \Big\{{x_1\over x_3}\Big\}}\Xi^{(a)}_{[1,1,0]}\right)
\ee

\paragraph{Young diagram [1,1,1].}

\be
E_{[1,1,1]}^{(a)}&=&\Xi^{(a)}_{[1,1,1]}
\ee

\paragraph{Young diagram [2,1].}

\be\label{21}
qtE_{[0,1,2]}^{(a)}&=&{\Big\{{qtx_3\over x_1}\Big\}\over \Big\{{qx_3\over x_1}\Big\}}\Xi^{(a)}_{[0,1,2]}+
{(1-t)\over(1-qt^2)}{\Big\{{qt^2x_1\over qx_3}\Big\}\over \Big\{{x_1\over qx_3}\Big\}}\Xi^{(a)}_{[1,1,1]}\nn\\
qt^2E_{[1,0,2]}^{(a)}&=&{\Big\{{qtx_3\over x_2}\Big\}\over \Big\{{qx_3\over x_2}\Big\}}
{\Big\{{tx_1\over x_2}\Big\}\over \Big\{{x_1\over x_2}\Big\}}\Xi^{(a)}_{[1,0,2]}+
{(1-t)\over(1-qt)}\left({\Big\{{tx_2\over x_3}\Big\}\over \Big\{{x_2\over qx_3}\Big\}}
{\Big\{{tx_1\over qx_3}\Big\}\over \Big\{{x_1\over qx_3}\Big\}}\Xi^{(a)}_{[1,1,1]}+
{\Big\{{qtx_3\over x_1}\Big\}\over \Big\{{qx_3\over x_1}\Big\}}
{\Big\{{qtx_2\over x_1}\Big\}\over \Big\{{x_2\over x_1}\Big\}}\Xi^{(a)}_{[0,1,2]}\right)\nn\\
qt^2E_{[0,2,1]}^{(a)}&=&{\Big\{{tx_2\over x_3}\Big\}\over \Big\{{x_2\over x_3}\Big\}}
{\Big\{{qtx_2\over x_1}\Big\}\over \Big\{{qx_2\over x_1}\Big\}}\Xi^{(a)}_{[0,2,1]}+
{(1-t)\over(1-qt)}\left({\Big\{{tx_1\over qx_3}\Big\}\over \Big\{{x_1\over qx_3}\Big\}}
{\Big\{{tx_1\over x_2}\Big\}\over \Big\{{x_1\over qx_2}\Big\}}\Xi^{(a)}_{[1,1,1]}+
{\Big\{{qtx_3\over x_1}\Big\}\over \Big\{{qx_3\over x_1}\Big\}}
{\Big\{{qtx_3\over x_2}\Big\}\over \Big\{{x_3\over x_2}\Big\}}\Xi^{(a)}_{[0,1,2]}\right)\nn\\
&\ldots&
\ee

\section{Properties of twisted non-symmetric Macdonald polynomials\label{PM}}

\subsection{Symmetry properties}

One can immediately obtain the stability property of the twisted non-symmetric Macdonald polynomials:
\be\label{stab}
\begin{array}{ll}
E^{(a)}_{[\alpha_1,\alpha_2,\ldots,\alpha_{n-1},\alpha_n]}(x_1,x_2,\ldots,x_{n-1},x_n)\Big|_{x_n=0}
=E^{(a)}_{[\alpha_1,\alpha_2,\ldots,\alpha_{n-1}]}(x_1,x_2,\ldots,x_{n-1})&\ \ \ \ \ \hbox{if}\ \ \ \alpha_n=0\cr
E^{(a)}_{[\alpha_1,\alpha_2,\ldots,\alpha_{n-1},\alpha_n]}(x_1,x_2,\ldots,x_{n-1},x_n)\Big|_{x_n=0}=0&\ \ \ \ \ \hbox{if}\ \ \ \alpha_n\ne 0
\end{array}
\ee

There is a symmetry
\be
E_{[0,1,1]}^{(a)}(x_1,x_2,x_3)&=&x_3^{1\over a}E_{[0,0,1]}^{(a)}(qx_3,x_1,x_2)\nn\\
E_{[1,0,1]}^{(a)}(x_1,x_2,x_3)&=&x_3^{1\over a}E_{[0,1,0]}^{(a)}(qx_3,x_1,x_2)\nn\\
E_{[0,0,2]}^{(a)}(x_1,x_2,x_3)&=&q^{-1}x_3^{1\over a}E^{(a)}_{[1,0,0]}(qx_3,x_1,x_2)\nn
\ee
These are particular cases of the general identity
\be\label{sym2}
\boxed{
E_{[\alpha_2,\ldots,\alpha_n,\alpha_1+1]}^{(a)}(x_1,x_2,\ldots,x_n)=
q^{-\alpha_1}x_n^{1\over a}E^{(a)}_{[\alpha_1\alpha_2,\ldots,\alpha_n]}(qx_{n},x_1,x_2,\ldots,x_{n-1})
}
\ee
which is a counterpart of (\ref{Bs}) in the twisted case. In order to derive it, we notice that, similarly to (\ref{B}), one has at generic $a$ from (\ref{BCa})
\be\label{Bac}
\boxed{
\mathfrak{B}^{(a)}\cdot E^{(a)}_{[\alpha_1,\ldots,\alpha_n]}(x_1,\ldots,x_n)=q^{1-a\over 2}t^{-\#\{\alpha_i\le\alpha_1\}}
E^{(a)}_{[\alpha_2,\ldots,\alpha_{n},\alpha_1+1]}(x_1,\ldots,x_n)
}
\ee
Acting on the both parts of this formula by $\mathfrak{C}^{(a)}_n$ and using (\ref{eva}), one comes to the symmetry property (\ref{sym2}).

\subsection{Permutations}

The action of operators $T_i$ just permutes the $i$-th and $(i+1)$-th parts of the \wc so that
\be\label{perm}
\boxed{\begin{array}{rcl}
T_iE^{(a)}_\alpha&=&E^{(a)}_\alpha,\ \ \ \ \ \ \hbox{if}\ \ \ \ \ \alpha_i=\alpha_{i+1}\cr\cr
T_iE^{(a)}_\alpha&=&C^{(1)}_{i,\alpha}E^{(a)}_\alpha+E^{(a)}_{\sigma_i\alpha},\ \ \ \ \ \ \hbox{if}\ \ \ \ \ \alpha_i<\alpha_{i+1}\cr\cr
T_iE^{(a)}_\alpha&=&C^{(1)}_{i,\alpha}E^{(a)}_\alpha+C^{(2)}_{i,\alpha} E^{(a)}_{\sigma_i\alpha},\ \ \ \ \ \ \hbox{if}\ \ \ \ \ \alpha_i>\alpha_{i+1}
\end{array}
}
\ee
where $\sigma_i\alpha$ permutes the $i$-th and $(i+1)$-th parts of $\alpha$, and $C^{(1)}_i$, $C^{(2)}_i$ are some constants of $q$ and $t$:
\be\label{C12}
C^{(1)}_{i,\alpha}:&=&-{(1-t)\Lambda^{(i,a)}_\alpha\over t(\Lambda^{(i,a)}_\alpha-\Lambda^{(i+1,a)}_\alpha)}=
-{(1-t)\Lambda^{(i)}_\alpha\over t(\Lambda^{(i)}_\alpha-\Lambda^{(i+1)}_\alpha)}=-{(1-t)\over t(1-r_{\alpha,i})}\\
C^{(2)}_{i,\alpha}:&=&{(\Lambda^{(i,a)}_\alpha-\Lambda^{(i+1,a)}_\alpha t)(\Lambda^{(i,a)}_\alpha-t^{-1}\Lambda^{(i+1,a)}_\alpha)\over t(\Lambda^{(i,a)}_\alpha-\Lambda^{(i+1,a)}_\alpha)^2}=
{(\Lambda^{(i)}_\alpha-\Lambda^{(i+1)}_\alpha t)(\Lambda^{(i)}_\alpha-t^{-1}\Lambda^{(i+1)}_\alpha)\over t(\Lambda^{(i)}_\alpha-\Lambda^{(i+1)}_\alpha)^2}={(1-tr_{\alpha,i})(1-t^{-1}r_{\alpha,i})\over t(1-r_{\alpha,i})^2}\nn
\ee
where we introduced the quantity
\be
r_{\alpha,i}:={\Lambda_\alpha^{(i+1)}\over\Lambda_\alpha^{(i)}}=q^{\alpha_{i+1}-\alpha_i}t^{\zeta(\alpha)_i-\zeta(\alpha)_{i+1}}\nn
\ee
Formulas (\ref{perm})-(\ref{C12}) follow from (\ref{TC2}). Since quantities (\ref{C12}) do not change when all the eigenvalues are rescaled at once, they do not depend on $a$ at all.

For instance:
\be
T_1E^{(a)}_{[0,0,1]}&=&E^{(a)}_{[0,0,1]}\nn\\
T_1E^{(a)}_{[0,1,0]}&=&-{1-t\over t(1-qt)}E^{(a)}_{[0,1,0]}+E^{(a)}_{[1,0,0]}\nn\\
T_1E^{(a)}_{[1,0,0]}&=&-{1-t\over t(1-q^{-1}t^{-1})}E^{(a)}_{[1,0,0]}+{(1-q)(1-qt^2)\over t(1-qt)^2}E^{(a)}_{[0,1,0]}\nn\\
T_2E^{(a)}_{[0,0,1]}&=&-{1-t\over t(1-qt^2)}E^{(a)}_{[0,0,1]}+E^{(a)}_{[0,1,0]}\nn\\
T_2E^{(a)}_{[0,1,0]}&=&-{1-t\over t(1-q^{-1}t^{-2})}E^{(a)}_{[0,1,0]}+{(1-qt)(1-qt^3)\over t(1-qt^2)^2}E^{(a)}_{[0,0,1]}\nn\\
T_2E^{(a)}_{[1,0,0]}&=&E^{(a)}_{[1,0,0]}\nn
\ee

\subsection{Symmetric twisted Macdonald polynomials}

Symmetric twisted Macdonald polynomials associated with the dominant integral weights can be obtained from the non-symmetric Macdonald polynomials by summing up over the Weyl group $W={\cal S}_n$, i.e. over all permutations of the partition $\lambda$:
\be\label{nss}
M_{\lambda}^{(a)}=\sum_{{\alpha=w\cdot\lambda}\atop{w\in W}} E^{(a)}_\alpha\cdot\left(\prod_{(i,j):\ {{\alpha_j>\alpha_i}\atop{j>i}}}{\Lambda^{(i)}_\alpha-t^{-1}\Lambda^{(j)}_\alpha\over
\Lambda^{(i)}_\alpha-\Lambda^{(j)}_\alpha}\right)
\ee
where $\lambda$ is a partition (Young diagram), the product in the summand runs over pairs of $(i,j)$ such that $\alpha_i<\alpha_j$ at $i<j$, and the sum runs over all permutations $w$ from the symmetric group $S_n$. These formulas follow from (\ref{perm}), and the coefficients are independent of $a$, as in (\ref{perm}), which {\bf proves the corresponding conjecture in \cite[Sec.6.4]{MMP1}}. Hence, they can be read off from the $a=1$ case when they give \cite{Mac96,MN} the symmetric Macdonald polynomials in the standard normalization of the $P$ polynomials \cite{Mac}.

\section{Algorithmic construction of twisted non-symmetric Macdonald polynomials}

\subsection{Algorithmic procedure}

Note that acting with operators $T_i$ and using (\ref{sym2}) (which we denote through ``B-operation" below) allows one to construct the twisted non-symmetric Macdonald polynomials recursively much similarly to how it is done in the non-twisted case \cite[Lemma 2.1.2]{HHL}.
The procedure consists of two steps, and is as follows.

\begin{itemize}
\item Assume one has all polynomials at level $|\alpha|-1$. Then, one picks up proper polynomials at that level in order to generate all possible $\alpha^-$'s at level $|\alpha|$ with the B-operation.
\item Then, one uses the $T_i$-operators in order to generate all \wcs from these $\alpha^-$'s.
\end{itemize}

For instance, in order to get the solution enumerated by the \wc $[0,0,3]$, one uses the following sequence:
\be
[0,0,0]\stackrel{B}{\longrightarrow}[0,0,1]\stackrel{T_2}{\longrightarrow}[0,1,0]\stackrel{T_1}{\longrightarrow}[1,0,0]
\stackrel{B}{\longrightarrow}[0,0,2]\stackrel{T_2}{\longrightarrow}[0,2,0]\stackrel{T_1}{\longrightarrow}[2,0,0]
\stackrel{B}{\longrightarrow}[0,0,3]\nn
\ee
Here by action of $T_i$ on $E^{(a)}_\alpha$ we mean that one has to subtract from $T_i(E^{(a)}_\alpha)$ the proper contribution of $E^{(a)}_\alpha$ in accordance with (\ref{perm}).

The action of B-operation is very simple. For instance,
\be
[0,0,0]\stackrel{B}{\longrightarrow}[0,0,1]\stackrel{B}{\longrightarrow}[0,1,1]\stackrel{B}{\longrightarrow}[1,1,1]
\stackrel{B}{\longrightarrow}[1,1,2]\stackrel{B}{\longrightarrow}\ldots\nn
\ee
corresponds to
\be
\Xi^{(a)}_{[0,0,0]}\longrightarrow\Xi^{(a)}_{[0,0,1]}\longrightarrow\Xi^{(a)}_{[0,1,1]}\longrightarrow\Xi^{(a)}_{[1,1,1]}\longrightarrow
\Xi^{(a)}_{[1,1,2]}\longrightarrow\ldots\nn
\ee
This procedure generates the \wc $\alpha^-$ maximally remote from the Young diagram. Other \wcs are obtained by moving parts of the \wc to the left by action of $T_i$'s, and each permutation increases the number of $\Xi^{(a)}_{\beta}$ emerging in expansion of $E^{(a)}_\alpha$,
and it increases the number of fractions in front of them by one, which {\bf proves the conjecture of \cite{MMP1} that the number of fractions $N_\alpha$ in $F_{\alpha,\beta}(\vec x)$ in formula (\ref{gE}) (which is the same for all $\beta$) is $N_{\alpha^-}$ plus the minimal length of permutation that brings the weak composition $\alpha$ to $\alpha^-$.}

On one hand, this procedure constructively {\bf proves the conjecture of \cite{MMP1} that the coefficients in front of $\Xi^{(a)}_{\beta}$ are rational functions that do not depend on $a$}. On the other hand, these coefficients are not that simple. Let us look at them in detail.

As soon as we need to move the larger numbers to the left, we need only the second line in (\ref{perm}). Then, we obtain for $E^{(a)}_{\sigma_i\alpha}$ using (\ref{gE}):
\be
E^{(a)}_{\sigma_i\alpha}=T_iE^{(a)}_\alpha-C^{(1)}_{i,\alpha}E^{(a)}_\alpha=
{1\over t}{\Big\{{tx_i\over x_{i+1}}\Big\}\over\Big\{{x_i\over x_{i+1}}\Big\}}\cdot\sigma_{i}E^{(a)}_\alpha
+\left(
{(1-t^{-1})\over\Big\{{x_i\over x_{i+1}}\Big\}}-C^{(1)}_{i,\alpha}\right)E^{(a)}_\alpha\nn
\ee
where $\sigma_i$ acts on $E^{(a)}_\alpha$ permuting $x_i$ and $x_{i+1}$.

Using (\ref{C12}), we obtain finally
\be\label{main}
\boxed{
E^{(a)}_{\sigma_i\alpha}=
{1\over t}{\Big\{{tx_i\over x_{i+1}}\Big\}\over\Big\{{x_i\over x_{i+1}}\Big\}}\cdot\sigma_{i}E^{(a)}_\alpha+{1\over t}
{(1-t)\over (1-r_{\alpha,i})}{\Big\{{r_{\alpha,i}x_{i+1}\over x_{i}}\Big\}\over \Big\{{x_{i+1}\over x_i}\Big\}}E^{(a)}_\alpha
}
\ee
or
\be\label{main1}
\boxed{
E^{(a)}_{\sigma_i\alpha}={1\over t}\sum_\beta
\left[{\Big\{{tx_i\over x_{i+1}}\Big\}\over
\Big\{{x_i\over x_{i+1}}\Big\}}\cdot\left[\sigma_i F_{\alpha,\beta}(\vec x)\right]\cdot \Xi^{(a)}_{\sigma_i\beta}+{(1-t)\over (1-r_{\alpha,i})}{\Big\{{r_{\alpha,i}x_{i+1}\over x_{i}}\Big\}\over \Big\{{x_{i+1}\over x_i}\Big\}}F_{\alpha,\beta}(\vec x)\cdot \Xi^{(a)}_{\beta}\right]
}
\ee
Note that this implies that the coefficients in front of products in $F_{\alpha,\beta}(\vec x)$ entering $F_{\lambda\mu}(x)$ are ratios of $q$-numbers (as was conjectured in \cite{MMP1}).

\subsection{Examples of algorithmic procedure}

For instance, acting by $T_1$ on $E^{(a)}_{[0,1,1]}$, in accordance with this formula, we obtain $E^{(a)}_{[1,0,1]}$:
\be
E^{(a)}_{[1,0,1]}=T_1E^{(a)}_{[0,1,1]}-\underbrace{C^{(1)}_{1,[0,1,1]}}_{{1-t^{-1}\over 1-qt^2}}E^{(a)}_{[0,1,1]}=
{1\over t}{\Big\{{tx_1\over x_2}\Big\}\over \Big\{{x_1\over x_2}\Big\}}\Xi^{(a)}_{[1,0,1]}
+{(1-t)\over t(1-qt^2)}{\Big\{{qt^2x_2\over x_1}\Big\}\over \Big\{{x_2\over x_1}\Big\}}\Xi^{(a)}_{[0,1,1]}\nn
\ee
which agrees with formula (\ref{main1}).

More interesting is the case when $\beta_i=\beta_{i+1}$. For instance, one obtains $E^{(a)}_{[0,2,0]}$ from $E^{(a)}_{[0,0,2]}$ by the action of $T_2$. $E^{(a)}_{[0,0,2]}$ contains the term with $\beta=[0,1,1]$:
\be
F_{[0,0,2],[0,1,1]}={(1-t)\over(1-qt)}{\Big\{{tx_2\over x_3}\Big\}\over \Big\{{x_2\over qx_3}\Big\}}{\Big\{{tx_2\over x_1}\Big\}\over\Big\{{x_2\over x_1}\Big\}}\nn
\ee
In accordance with (\ref{main1}), it results in two terms in $F_{[0,2,0],[0,1,1]}$, i.e. in front of $\Xi^{(a)}_{[0,1,1]}$ in $E^{(a)}_{\sigma_2[0,0,2]}$:
\be\label{2t}
{(1-t)\over (1-qt)}\left({\Big\{{tx_2\over x_3}\Big\}\over \Big\{{x_2\over x_3}\Big\}}{\Big\{{tx_3\over x_2}\Big\}\over \Big\{{x_3\over qx_2}\Big\}}{\Big\{{tx_3\over x_1}\Big\}\over \Big\{{x_3\over x_1}\Big\}}+{(1-t)\over (1-q^2t^2)}
{\Big\{{q^2t^2x_3\over x_2}\Big\}\over \Big\{{x_3\over x_2}\Big\}}{\Big\{{tx_2\over x_3}\Big\}\over \Big\{{x_2\over qx_3}\Big\}}
{\Big\{{tx_2\over x_1}\Big\}\over \Big\{{x_2\over x_1}\Big\}}\right)
\ee
This combination is looking different from that in \cite[Eq.(107)]{MMP1}, but is actually the same.

In order to generate further $E^{(a)}_{[2,0,0]}=E^{(a)}_{\sigma_1[0,2,0]}$, one acts by $T_1$ on $E^{(a)}_{[0,2,0]}$. There is a conspiracy there. For instance, the coefficient in front of $\Xi^{(a)}_{[0,1,1]}$ in $E^{(a)}_{\sigma_2[2,0,0]}$ comes from two terms (\ref{2t}) and also from the coefficient in front of $\Xi^{(a)}_{\sigma_1[1,0,1]}$. The sum of these three terms miraculously produces a single term in front of $\Xi^{(a)}_{[0,1,1]}$ in $E^{(a)}_{[2,0,0]}$, (\ref{l2}):
\be
F_{[2,0,0],[0,1,1]}={(1-t)\over (1-qt)}\left({\Big\{{tx_2\over x_3}\Big\}\over \Big\{{x_2\over x_3}\Big\}}{\Big\{{tx_3\over x_2}\Big\}\over \Big\{{x_3\over qx_2}\Big\}}{\Big\{{tx_3\over x_1}\Big\}\over \Big\{{x_3\over x_1}\Big\}}+{(1-t)\over (1-q^2t^2)}
{\Big\{{q^2t^2x_3\over x_2}\Big\}\over \Big\{{x_3\over x_2}\Big\}}{\Big\{{tx_2\over x_3}\Big\}\over \Big\{{x_2\over qx_3}\Big\}}
{\Big\{{tx_2\over x_1}\Big\}\over \Big\{{x_2\over x_1}\Big\}}\right)\times {(1-t)\over(1-q^2t)}
{\Big\{{q^2tx_2\over x_1}\Big\}\over \Big\{{x_2\over x_1}\Big\}}+\nn\\
+{(1-t^2)\over(1-qt)(1-q^2t^2)}{\Big\{{tx_1\over x_2}\Big\}\over \Big\{{x_1\over x_2}\Big\}}
{\Big\{{q^2t^2x_3\over x_1}\Big\}\over \Big\{{x_2\over x_1}\Big\}}{\Big\{{tx_2\over x_3}\Big\}\over \Big\{{x_2\over qx_3}\Big\}}
{\Big\{{tx_2\over x_1}\Big\}\over \Big\{{x_2\over x_1}\Big\}}=
{q(1+q)(1-t)^2\over(1-qt)(1-q^2t)}
{\Big\{{tx_3\over x_2}\Big\}\over \Big\{{qx_3\over x_2}\Big\}}{\Big\{{tx_2\over x_3}\Big\}\over\Big\{{qx_2\over x_3}\Big\}}{\Big\{{qtx_3\over x_1}\Big\}\over \Big\{{x_3\over x_1}\Big\}}{\Big\{{qtx_2\over x_1}\Big\}\over \Big\{{x_2\over x_1}\Big\}}\nn
\ee

\section{Conclusion}

\subsection{Summary}
Let us briefly summarize the emerging pattern of eigenfunctions of the twisted Cherednik Hamiltonians described in the previous sections.

\begin{itemize}
\item{} The eigenfunctions produce a system of excitations over the ground state.
\item{} The ground state eigenfunction $\Omega^{(a)}_m(\vec x)$ of the ground state is still a rather complicated function,
which is explicitly known only in particular cases.
\item{} Excitations $E_\alpha^{(a)}(\vec x)$
instead are reasonably simple,
they depend on the twist $a$ only through $\Omega^{(a)}_m(\vec x)$.
\begin{itemize}
\item[$\diamond$] They are labeled by weak compositions $\alpha=\{\alpha_1,\ldots,\alpha_n\}$, $\alpha=s(\lambda)$ being an arbitrary permutation $s$
of lines in the Young diagram (partition) $\lambda=\{\lambda_1,\ldots,\lambda_n\}$ (with $\lambda_1\ge\lambda_2\ge\ldots\ge\lambda_n\ge 0$).
\item[$\diamond$] At $t=q^{-m}$, the eigenfunctions are polynomials of $\vec x^{1\over a}$, decomposed into peculiar sums
\be
E^{(a)}_{\alpha}(\vec x)=\sum_{\beta\le\alpha} F_{\alpha,\beta}(\vec x)\cdot \Xi^{(a)}_{\beta}
\label{summaryEdeco}
\ee
where $\Xi^{(a)}_\beta$ are directly made from $\Omega^{(a)}_m$,
and coefficients $F_{\alpha\beta}^{(m)}$ are {\it independent} of the twisting $a$
and are {\it rational}(!) functions of $\vec x$, combined into polynomials due to peculiar
identities for the polynomials $\Omega^{(a)}_m$, which have not yet been tamed in any systematic way.
\item[$\diamond$] The eigenfunctions can be constructed by action of the permutation operators $T_i$ producing
$E_{\sigma_i\alpha}$ from $E_{\alpha}$,
and of the creating operators $\mathfrak{B}^{(a)}$, producing $E_{\alpha}$ from the lower level $E_\alpha'$, $|\alpha|=|\alpha'|+1$ by adding a box to the weak composition, which is a direct generalization of the Knop-Sahi recursion \cite{KS,HHL}.
\item[$\diamond$] Technically effective is first to build the eigenfunctions $E_{\alpha^-}$ with $0\le\alpha_n\le\alpha_{n-1}\le\ldots\le\alpha_1$ by the creation operators and then to use the permutation operators $T_i$ in order to generate all other $E_{\alpha}$ with the same $\alpha^-$.
\item[$\diamond$] In fact, there are many ways to reach a given $\alpha$ starting from the ground state $\{0,\ldots,0\}$, but, nevertheless, they produce unambiguous result.
The system of relations between these different combinations of creation and permutation operators that ensures it is not yet studied in detail.
\item[$\diamond$] There is not yet any clear reason for the coefficients $F_{\alpha,\beta}^{(m)}(\vec x)$
to be factorized: the permutation operators produce them in the form of long sums.
But in the so-far studied examples the sums become very short if exist at all.
\end{itemize}
\item{} Due to the independence of $a$, many features of the coefficients $F_{\alpha,\beta}^{(m)}$
can already be observed for non-twisted non-symmetric Macdonald polynomials (at $a=1$)
and even for non-symmetric Jack polynomials.
However, at $a=1$, there is no clear way to see the very need of decomposing polynomials into combinations of rational functions
like (\ref{summaryEdeco}), nor a way to unambiguously do so, this explains why this essential structure has not been well-known for researchers so far.
\end{itemize}

\subsection{Concluding remarks}

To conclude, in this paper, we proved the three conjectures that we proposed in \cite{MMP1}:
\begin{itemize}
\item The coefficients $F_{\alpha,\beta}(\vec x)$ in front of $\Xi^{(a)}_{\beta}$ in formula (\ref{gE}) are rational functions that do not depend on $a$.
\item The number of fractions in $F_{\alpha,\beta}(\vec x)$ is equal to the minimal length of permutation that brings the weak composition $\alpha$ to $\alpha^-$.
\item Symmetric polynomials are made from $E^{(a)}_{\alpha}$ by formula (\ref{nss}).
\end{itemize}

We have built an algorithmic procedure that allows one to generate arbitrary twisted (non-polynomial) eigenfunctions basing on the ground state solution $\Omega^{(a)}(\vec x)$. However, this solution is known only in the case of $t=q^{-m}$ with natural $m$ due to its relation with the twisted Baker-Akhiezer function (\ref{omegaBA}). Continuation to an arbitrary $t$ is generally unknown. It can be sometimes done in the case when one has an explicit formula at hands. For instance, in the simplest twisted $n=2$, $a=2$ case, formula (\ref{omega22}) can be directly extended \cite{MMP1} to
\be
\Omega^{(2)}(x_1,x_2;q,t)=t^{-{1\over 2}\log_q x_1}{\Big(-\sqrt{qtx_2\over x_1};q\Big)_\infty
\over \Big(-\sqrt{qx_2\over tx_1};q\Big)_\infty}\nn
\ee
However, even having explicit sums in the case of arbitrary twist $a$ at $n=2$ (see \cite[Eq.(94)]{MMP1}), it is not that clear how to extend them to an arbitrary $t$.
Generally, one needs a counterpart of the Noumi-Shiraishi power series \cite{NS} in the twisted case, i.e. a twisted counterpart of the triad \cite{MMPBA3}, which is unknown yet.

Another unclear point is the conspiracy illustrated at the very end of the previous section: typically the coefficients in front of each $\Xi^{(a)}_\alpha$, which are sums of many terms, either factorize, or at least can be presented as a sum of a much smaller number of factorized terms. The mechanism behind this phenomenon and the overall answer for these coefficients remain unclear.

It should be noted that {\bf symmetric} Macdonald polynomials are known to be constructed from a trivial polynomial by applying proper creation operators \cite{KN,MMkn}. It is natural to ask what are counterparts of these operators in the twisted case.

We are planning to return to these problems elsewhere.

\section*{Acknowledgements}

The work was partially funded within the state assignment of the NRC Kurchatov Institute, was partly supported by the grant of the Foundation for the Advancement of Theoretical Physics and Mathematics “BASIS” and by Armenian SCS grants 24WS-1C031.

\newpage

\section*{Appendix}

We attach to this submission a MAPLE file that allows one to generate the twisted and non-twisted Macdonald polynomials, both non-symmetric and symmetric. In fact, the non-twisted Macdonald polynomials are certainly available with much more effective Maple packages \cite{SF}.
Here they remain as a special case of twisted polynomials, since our code is designed to work with twisted polynomials. The twisted polynomials are expressed through the ground state $\Omega^{(a)}(\vec x)$ in the code, and the only difficult point remaining here is an explicit expression for $\Omega^{(a)}(\vec x)$. At the moment, there is a closed formula for this function only at $n=2$, any $a$ in \cite[Eqs.(74),(94)]{MMP1}, and it is listed for other particular small values of parameters in the file Omega3.txt attached to the arXiv version of \cite{MMP2}.

\bigskip

Here is the example of how to use the MAPLE file.

First of all load the file with simple \texttt{read} command
\begin{flalign}
  \texttt{read("md-calc.mpl");} &\nn
\end{flalign}

The usual symmetric Macdonald polynomials are given by the \texttt{mac} function:
\begin{flalign}
  \texttt{mac([2,0]);} & &\nn
  \\ \notag & \mathcal{M}_{[2,0]}(x_1,x_2) = x_1^2 + \frac{(q+1)(t-1)}{(q t-1)} x_1 x_2 + x_2^2
\end{flalign}

The non-symmetric counterparts of usual Macdonald polynomials are given by the
\texttt{ns\_mac} function:
\begin{flalign}
  \texttt{ns\_mac([2,0]);} & &\nn
  \\ \notag & E_{[2,0]}^{(1)}(x_1,x_2) =
  x_1^2+\frac{q (t-1) (q+1)}{(q^2 t-1)} x_1 x_2+\frac{(t-1) q^2}{(q^2 t-1)} x_2^2
\end{flalign}

The non-symmetric twisted Macdonald polynomials, the main subject of this paper,
are calculated with help of the \texttt{twisted\_ns\_mac} function
\newcommand\bsb[1]{
  \Big\{#1\Big\}
}
\newcommand\xto[0]{
  {x_2\over x_1}
}
\be
  \texttt{twisted\_ns\_mac([2,0]);} & &\nn
  \\ qt^2E_{[2,0]}^{(a)} &=&
{\Big\{{qtx_1\over x_2}\Big\}\over\Big\{{qx_1\over x_2}\Big\}}{\Big\{{tx_1\over x_2}\Big\}\over \Big\{{x_1\over x_2}\Big\}}\Xi^{(a)}_{[2,0]}+
{(1-t)\over(1-q^2t)}
  {\Big\{{qtx_2\over x_1}\Big\}\over \Big\{{qx_2\over x_1}\Big\}}{\Big\{{q^2tx_2\over x_1}\Big\}\over \Big\{{x_2\over x_1}\Big\}}\Xi^{(a)}_{[0,2]}+\nn\\
&&+{q(1+q)(1-t)\over (1-q^2t)}{\Big\{{tx_1\over x_2}\Big\}\over\Big\{{qx_1\over x_2}\Big\}}{\Big\{{qtx_2\over x_1}\Big\}\over \Big\{{qx_2\over x_1}\Big\}}\Xi^{(a)}_{[1,1]}\nn
\ee
and, similarly, the \texttt{twisted\_mac} function gives the symmetric twisted Macdonald
polynomials, according to \eqref{nss}.

\bigskip

Of note, there are also helper functions \texttt{slurp\_xi} and \texttt{deslurp\_xi}
which switch between hiding the $x$-rescaling inside $\Xi$-functions and manifestly
writing the rescaled $\Omega(\vec{x})$, and the underlying \texttt{twisted\_mac1}
function, of which all the above \texttt{mac}-functions are just the convenience wrappers.

\end{document}